\documentstyle[epsf,aps,prl,preprint]{revtex}

\def\rep#1{${\bf #1}$}
\def\repC#1{${\bf\overline{#1}}$}
\def\be{\begin{equation}}
\def\ee{\end{equation}}
\def\bea{\begin{eqnarray}}
\def\eea{\end{eqnarray}}

\def\Mz{M_Z}
\def\Msusy{M_{\rm SUSY}}
\def\Mgut{M_{\rm GUT}}
\def\Mpd{M_{\rm PD}}
\def\GeV{\; {\rm GeV}}
\def\yGUT{y_{\rm GUT}}

\begin{document}
\draft

\preprint{\vbox{
\hbox{CTP-TAMU-60/96}
\hbox{hep-ph/9611389}
\hbox{November 1996}
}}

\title{Proton Decay and the Dimopoulos-Wilczek Mechanism in Minimal
       SO(10) Models}
\author{S.~Urano\cite{shinichi} and R.~Arnowitt\cite{dick}}

\address{Center for Theoretical Physics, Department of Physics,
	 Texas A\&M University, College Station, TX 77843-4242}

\maketitle

\begin{abstract}

Proton decay is examined within the framework of certain SO(10) 
models which have low dimension representations 
compatible with the recently found constraints coming from
string theory, and which use the Dimopoulos-Wilczek mechanism to solve the
doublet-triplet splitting problem.  
It is found that the mass parameter that controls proton decay rate is
intimately related to the low energy data via the coupling unification
condition, and that a suppression of proton decay is achieved at the cost
of large threshold effects.
Furthermore, in cases where there are states with intermediate masses,
the threshold effects further enhance proton decay.
The experimental bound then severely constrains the parameter space
of these models.  Some possible solutions are suggested.

\end{abstract}
\narrowtext

SO(10) is an attractive gauge group for grand unification.  It may be thought
of as arising via gauging the global B-L symmetry of SU(5) models.  In SO(10)
models, all the quarks and leptons for each generation are unified into a 
single representation, \rep{16}, and since SO(10) observes
left-right symmetry, anomaly cancellation is automatic.
Also, the Yukawa couplings for the members of a family
(e.g., top, bottom, and $\tau$ for the third generation) must be the same
at the GUT scale and this has allowed a successful
prediction for the mass of the top quark, provided that the value of
$\tan \beta$ is large, $\sim 50$.  Another nice feature of SO(10) 
is the existence of a way to avoid
the doublet-triplet splitting problem via the so called
``Dimopoulos-Wilczek mechanism.''\cite{Dimopoulos:1981inp,Srednicki:1982}

A drawback to SO(10) models in the past has been the largeness of the
particle content, due to the use of large representations such as \rep{126}
and \rep{210}.
From the viewpoint that a grand unified theory (GUT) is an
effective field theory of some Planck scale theory which is perturbative
below the Planck scale (e.g., a string model), it is desirable
not to have a large number of chiral multiplets since these then would drive
the unified coupling constant upwards, exceeding the perturbative bound before
reaching the Planck scale.  This desire is consistent with recent results
in free field heterotic string models\cite{Dienes:1996a}, where it has 
been shown that no massless SO(10) representations larger than \rep{54}
can arise, regardless of the affine level at which SO(10) is realized.

There are now several SO(10) models where the Dimopoulos-Wilczek mechanism
is employed while using only small particle 
content\cite{Hisano:1994,Babu:1995}.  Here, we will refer
to these as ``minimal'' SO(10) models.
In this letter, we present two new results that are relevant to proton
decay in these minimal SO(10) models, 
gotten by imposing the coupling constant unification condition.
First we show that,
contrary to what is found in the current literature\cite{foot1}, having a
mixing via the minimal $2 \times 2$ mass matrix among the color triplet Higgses 
does not of itself give rise to a new degree of freedom which can be 
varied in order suppress the proton decay rate.
Second, we show that the intermediate ($\sim 10^{10}$ GeV) mass states
which often arise in these models, coming from weakly coupling the two 
gauge breaking sectors, leads to an enhancement of proton decay rate.
These results, combined with 
the fact that proton decay is already enhanced considerably in SO(10) models
due to the largeness of $\tan \beta$, pushes the 
values of the parameters of these models into unnatural ranges when the 
experimental bound on proton decay is taken into account.  Below, we will
demonstrate these points within a general framework.
We will show elsewhere a more explicit demonstration of these points using the 
specific model by Babu and Barr\cite{Babu:1995}.

First we consider proton decay.   In minimal SO(10) models, there are two
\rep{10}'s of Higgses, each of which decomposes under SU(5) as 
\rep{5} + \repC{5}.  Generally, if there is more than one 
\rep{5} + \repC{5} pair, there is a possibility of mixing between the
triplet partners of the light doublets, that couple to the quarks and leptons, 
via a mass matrix.  When such a mixing occurs, the superpotential
may be written schematically as 
$\bar{H}_1 J + \bar{K} H_1 + {\cal M}_{IJ} \bar{H}_I H_J$,
where the \rep{5}+\repC{5} pairs are denoted by the indices ($I,J$),
and we have chosen a basis where $H_1$ and $\bar{H}_1$
are the Higgs which couple to the quarks and leptons.
The quark and lepton currents are denoted by $J$ and $\bar{K}$ and
${\cal M}$ is the mass matrix.
Integrating out $H_1$ then gives 
$ -\bar{K} ({\cal M}^{-1})_{11} J.$
Using a somewhat less bulky notation, $1/\Mpd \equiv ({\cal M}^{-1})_{11}$,
we then find two baryon number violating dimension four terms in the
superpotential suppressed by $\Mpd$, identical to the usual minimal SU(5)
case with the mass of the color triplet, $M_{H_3}$, replaced with $\Mpd$.

The actual calculation to
estimate the proton decay rate is quite complicated and will not
be repeated here.
An extensive calculation was presented
in \cite{Nath:1985,Arnowitt:1994}, where the $p \rightarrow \bar{\nu} K^+$
decay rate was given by
$ \Gamma(p \rightarrow \bar{\nu}_i K^+) = C (\beta_p/\Mpd)^2 |AB_i|^2.$
Here, $C$ and $\beta_p$ depend upon strong interaction parameters,
while $A$ depends on electroweak ones.
The quantity $B_i$ represents the chargino-squark-squark
triangle dressing loop, and includes the SUSY mass
dependences.  We note here that there is a $\sin 2 \beta$ factor
in its denominator.  Thus, the the decay rate is proportional to
$1/(\Mpd \sin(2\beta))^2$ which, for large $\tan \beta$,
is $\sim (\frac{\tan\beta}{\Mpd})^2$.
An analysis, running over the SUSY mass parameter space gives,
from Kamiokande data,
$\Mpd > 1.2 \times 10^{16}$~GeV
where the bound involves a lower bound of
$\tan \beta$ to be $\sim 1.4$\cite{Arnowitt:1994} which comes
from the top Yukawa developing a Landau pole at the GUT scale.
Noting the $\tan \beta$ dependency in the decay rate,
we then require
\be
\Mpd > \tan \beta \times 0.57 \times 10^{16} \ \GeV.
\label{eq:mpdlim}
\ee

Since one expects $\Mpd$ to be of order of the GUT scale,
$\sim 2.4 \times 10^{16}$ GeV, this bound poses a serious
problem to SO(10) models for which $\tan \beta$ is expected to be
$O(50)$, i.e., one must arrange to have $\Mpd$ significantly
larger than the GUT scale.

In the models we are studying, the Dimopoulos-Wilczek (DW) mechanism
is used to create the doublet-triplet splitting.  The basic idea is
to achieve a breaking of SO(10) via growing a VEV for a \rep{45}
of the form: 
$\langle A \rangle = {\rm diag }(a,a,a,0,0) \otimes i \sigma_2 $
where A is the \rep{45} in the 10 $\times$ 10 anti-symmetric tensor
notation.  This VEV 
breaks SO(10) down to the SU(3)$\times$SU(2)$\times$U(1)$\times$U(1) 
subgroup where the extra U(1) is identified with the
B$-$L symmetry.  The extra U(1) is broken in minimal models via VEV
growth for a \rep{16}+\repC{16} pair.  In the one-step unification
picture, we would expect $a$ to be $O(\Mgut)$.

If we now couple the \rep{45} to a pair of \rep{10}'s in a natural way and
have a mass term only for one of the \rep{10}'s, then the superpotential
will be
$ W_H = M_2 H_2^2 + \Lambda H_1 A H_2 $
where $H$'s are the \rep{10}'s.  Substituting in the VEV for $A$,
the mass terms for the doublet parts (indicated by superscript $d$)
and the triplet parts (indicated by superscript $t$) are then
\be
\begin{array}{c} 
\left( \begin{array}{cc} \bar{H_1^d} & \bar{H_2^d} \end{array} \right) \\ ~ 
\end{array}
\left( \begin{array}{cc} 0 & 0 \\ 0 & M_2 \end{array} \right)
\left( \begin{array}{c} H_1^d \\ H_2^d \end{array} \right)
\label{eq:massMatHd}
\ee
and
\be
\begin{array}{c}
\left( \begin{array}{cc} \bar{H_1^t} & \bar{H_2^t} \end{array} \right) \\ ~
\end{array}
\left( \begin{array}{cc} 0 &  \Lambda a \\ \Lambda a & M_2 \end{array} \right)
\left( \begin{array}{c} H_1^t \\ H_2^t \end{array} \right).
\label{eq:massMatHt}
\ee
Here the bar in the $\bar{H}$ indicate that it is part of the
\repC{5} under the SU(5) decomposition, \rep{10}=\rep{5}+\repC{5}.
Thus one of the doublets is automatically massless.  
These mass matrices also suggest the possibility of suppressing
proton decay by adjusting the parameters $\Lambda$ and $M_2$.  The
massive parameter which controls Higgsino-mediated proton decay is given by
\be
1/\Mpd \equiv \left( {\cal M}^{-1} \right)_{11} = M_2 / (\Lambda^2 a^2),
\label{eq:DWmPD}
\ee
where ${\cal M}$ is the triplet mass matrix.  By choosing $M_2$ to be
smaller than $\Lambda a$ (expected to be $O(\Mgut)$), it appears possible
to adjust $\Mpd$ so that it is significantly larger than $\Mgut$.
However since these parameters are also correlated to the mass spectrum
of the model, which affect the coupling constant running, we must check
if this suppression does not ruin the coupling unification picture.  This
is the task we now turn to.

In order to discuss the running of the coupling constants, it is
convenient to introduce the variables:
$y_i \equiv 1/\alpha_i \equiv 4 \pi/g_i^2$ and 
$x \equiv \frac{1}{2\pi} \ln(\mu/\Mz)$.
Then the RGE's ( $\mu \frac{\partial g_i}{\partial \mu} = \beta_i $)
can be written, to two-loop order, as
\be
\frac{d y_i}{d x} =
b_i + \frac{1}{4\pi} \sum_j b_{ij} \frac{1}{y_j},
\label{eq:yRGE2}
\ee
where the $b_i$ are the one-loop $\beta$ coefficient and $b_{ij}$
are the two-loop coefficients.

The one-loop solutions are just straight lines in the $x$-$y$ plane with
the slopes given by the one-loop coefficient.  
The thresholds corrections in mass-independent
renormalization schemes are then just a change
in the slope at the corresponding $x$'s.
The $\beta$ coefficients for arbitrary gauge theories may be found
in \cite{Jones:1982}.  

The RGE's can not be solved exactly at two-loop,
but the solution can be approximated by iterating
the one-loop result.  The error introduced in this approximation is
then expected to be the size of the next-order correction.
The threshold effects at two-loop are expected to be small.  In
the numerical results below, we have considered it in two situations: 
(1) when the
threshold involves many particles so that the change in the two-loop
$\beta$ coefficients is large, and (2) when the threshold involves
a large hierarchy in the energy scales
(i.e. when there exists an intermediate scale).

The condition for unification can be stated in the $\overline{\rm DR}$ scheme
as simply the coming together of the coupling constants to one
point\cite{Antoniadis:1982}.
The threshold corrections would then make the running of all the coupling
constants the same.
Calling the running coupling constant in the GUT to be $\yGUT$,
the unification condition is then
$\yGUT(x) = y_i(x), \;\;\; x \ge x_U$,
where $x_U$ is the highest threshold.
Adding in the threshold corrections at the one-loop level can be done
with just simple analytical geometry, and we arrive at
\be
\yGUT(x) = y_i(0) + b_{\rm GUT} x + \delta y^2_i
               - \sum_a b_{ai} x_a,
\label{eq:unify}
\ee
where $\delta{y}^2_i$ denotes the two-loop level contribution,
$y_i(0)$ are the inverses of the coupling constants measured
at $\Mz$, $b_{\rm GUT}$ is the one-loop 
$\beta$ coefficient in the unbroken GUT, and
$x_a \equiv \frac{1}{2\pi} \ln(\frac{M_a}{\Mz})$ where
$M_a$'s are the threshold masses.

Eq.~\ref{eq:unify} represents three constraints on the parameters of
a GUT model.  Of these, we always need to assign one to solve for $\yGUT$
whose experimental constraint is less stringent than the constraint we can
impose on it by grand unification.  Thus there are two constraints
upon the remaining parameters of a model.  
We can describe these constraints in terms of three-dimensional
vectors which we may dot into the unification condition.
It turns out, from the study of $\beta$ coefficients, that the two useful
vectors are $(1,-3,2)$ and $(-5,3,2)$, where the first one 
is relevant for $\Mpd$.  

\begin{figure}
\epsfxsize=0.9 \columnwidth
\epsffile{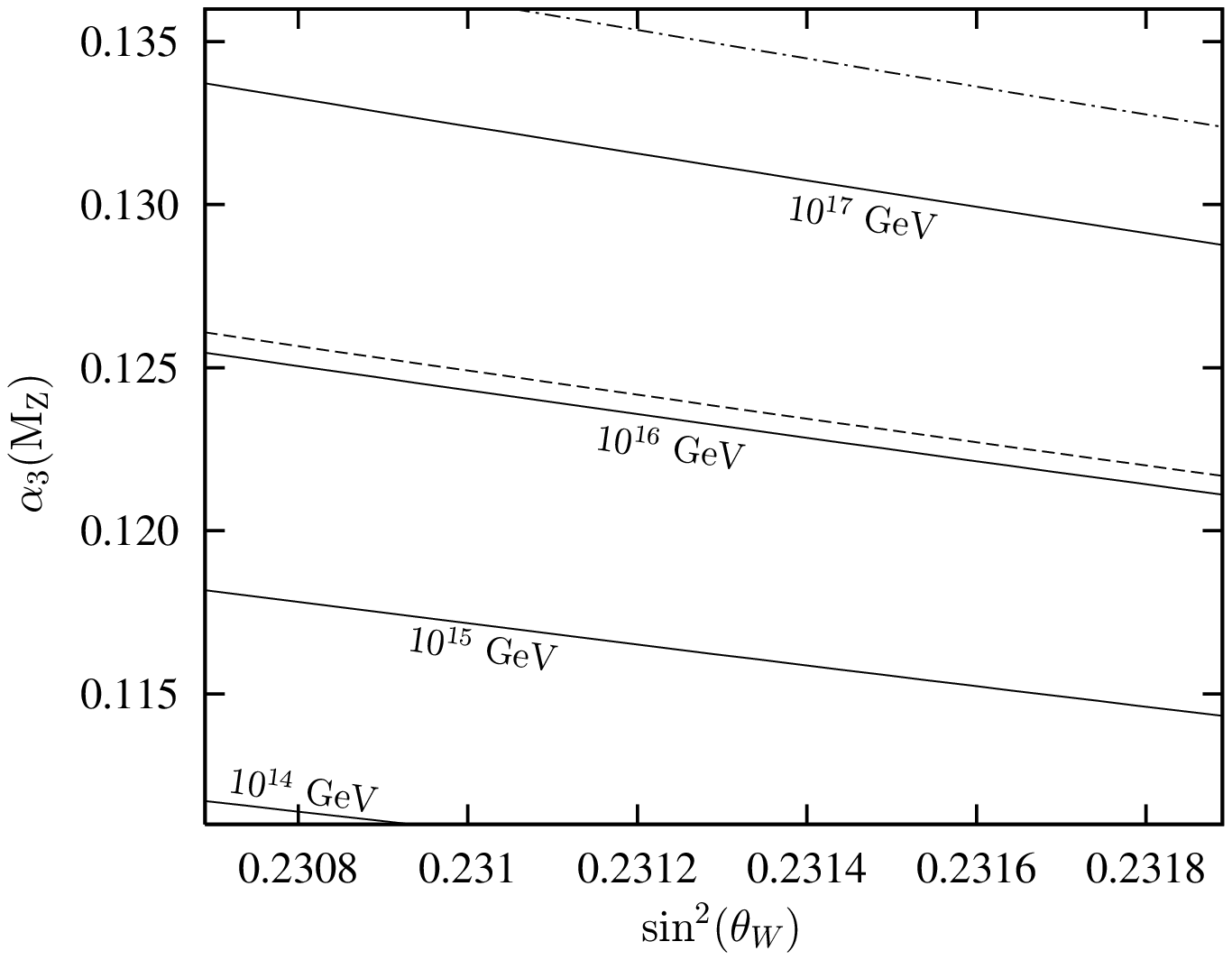}
\medskip
\caption{Contour Plot of $M_{H_3}^{(0)}$ in $\sin^2(\theta_W)$--$\alpha_3(\Mz)$
plane.  The dashed line corresponds to the lower bound on $\Mpd$ in minimal 
SU(5) of $1.2 \times 10^{16}$~GeV.  The dot-dashed line corresponds to the
lower bound on $\Mpd$ in minimal SO(10) of $2.7 \times 10^{17}$~GeV.  Currently,
the measurements are $\sin^2(\theta_W) = 0.2313 \pm 0.0003$ and 
$\alpha_3(\Mz) = 0.119 \pm 0.003$\protect\cite{Langacker:dpf96}.}
\label{fig1}
\end{figure}

We consider first the minimal SU(5) model, the superheavy particles are
the ((\rep{2},\rep{3})$_{-5/3}$ + h.c.) components of SU(5) gauge fields
(vector multiplet) in the \rep{24} representation ($V$), the
((\rep{3},\rep{1})$_0$ + (\rep{1},\rep{8})$_0$) 
components of the Higgs in the \rep{24} ($A$),
and the color triplet Higgs 
((\rep{1},\rep{3})$_{-2/3}$ + h.c.) of the Higgs in
the (\rep{5}+\repC{5}) ($H$).  We have used the notation 
(SU(2) rep., SU(3) rep.)$_{\rm Hypercharge}$ to show the representations under 
the Standard Model gauge group.
The three superheavy threshold masses are labeled $M_V$, $M_A$, and
$M_{H_3}$.  The proton decay mass parameter, $\Mpd$ is here just $M_{H_3}$.
Dotting the $(1,-3,2)$ vector into the unification condition, 
Eq.~\ref{eq:unify}, yields the necessary condition for $\Mpd$:
\be
\Bigl( \frac{M_{H_3}}{M_{H_3}^{(0)}} \Bigr) =
\Bigl( \frac{\Mpd}{M_{H_3}^{(0)}} \Bigr) =
\Bigl( \frac{ \Msusy }{ \Mz } \Bigr)^{5/6}
\label{eq:MpdBB0}
\ee
where $M_{H_3}^{(0)}$ is a quantity which is determined mainly by
the low energy measurement of the coupling constant, i.e.\ it is 
what the mass of the Higgs triplet would be for minimal SU(5) when
SUSY threshold corrections are neglected ($\Msusy = \Mz$).  It
can be determined by:
\bea
\frac{1}{2\pi} \ln \frac{M_{H_3}^{(0)}}{\Mz}
&=& \frac{1}{4 \alpha(\Mz)} (6 \sin^2(\theta_W) - 1) - \frac{5}{6 \alpha_3(\Mz)}
 \nonumber \\
&& + \frac{1}{2} \sum_i (1,-3,2)_i \delta{y}^2_i.
\eea
This quantity is plotted in Fig.~\ref{fig1} 
in the $\sin^2(\theta_W) - \alpha_3(\Mz)$ plane, using the estimated
two-loop correction by setting the GUT scale to be $\Mpd = M_{H_3}$.  
(We have used $\alpha(\Mz) = 1/127.9$.)
The right hand side of Eq.~(\ref{eq:MpdBB0}) 
is not expected to be large, i.e.\ $O(1)$, and thus
Fig.~\ref{fig1} may be viewed as bounds on $M_{H_3}$ as a function of
$\sin^2(\theta_W)$ and $\alpha_3(\Mz)$.
The proton decay constraint of
$M_{H_3} > 1.2 \times 10^{16}$~GeV 
(for the minimal SU(5), i.e. small $\tan \beta$) indicated by the dotted line
can be seen to push the satisfactory
$\alpha_3(\Mz)$ and $\sin^2(\theta_W)$ values
to the corner of the experimentally allowed region.  
This is in agreement with the well-known result that it is
difficult to achieve low values of $\alpha_3(\Mz)$ in the minimal SU(5) 
model\cite{Bagger:1995,Ring:1995}.

We are interested in obtaining a similar restriction on $\Mpd$ for SO(10).
For the models with two \rep{10}'s where the doublet-triplet splitting
is done by DW mechanism, we have seen that $\Mpd = (\Lambda^2 a^2)/M_2$
(Eq.~\ref{eq:DWmPD}).  From the mass matrices in Eqs.~\ref{eq:massMatHd}
and \ref{eq:massMatHt}, we see that $\Mpd$ is the product of the
triplet masses eigenstates divided by the heavy doublet mass.  Or,
using $T_3$ and $H_3$ to denote the color triplet eigenstates, and
$T_2$ to denote the heavy doublet eigenstate,
$ \Mpd = M_{T_3} M_{H_3} / M_{T_2}. $
This turns out to be the same combination of the threshold masses which
appear in the unification condition gotten by using the vector $(1,-3,2)$.
Neglecting for the moment 
the threshold effects due to the splittings among other superheavy
particles, we find that the analogue of Eq.~\ref{eq:MpdBB0} is:
\be
\Bigl( \frac{M_{H_3}}{M_{H_3}^{(0)}} \Bigr) =
\Bigl( \frac{ \Msusy }{ \Mz } \Bigr)^{5/6}
\Bigl( \frac{M_{T_2}}{M_{T_3}} \Bigr),
\label{eq:Mpd1}
\ee
or,
\be
\Bigl( \frac{\Mpd}{M_{H_3}^{(0)}} \Bigr) =
\Bigl( \frac{ \Msusy }{ \Mz } \Bigr)^{5/6},
\label{eq:Mpd2}
\ee
which is exactly the same as in the minimal SU(5) case.  
Thus, having a DW derived mass matrix mixing does not offer an 
advantage over the minimal SU(5) case in suppressing proton decay rate
since it is precisely the same combination of parameters which affects
proton decay that is constrained by the unification condition.

A loosening of the constraint on $\Mpd$
can come from threshold effects arising
from the splittings amongst other superheavy particles in a model.
In general, minimal SO(10) models contain
vector multiplets in the adjoint (\rep{45}) representation, plus
chiral multiplets in \rep{45}, \rep{54}, \rep{16}
+ \repC{16}, \rep{10}, and singlet representations.
Upon breaking to the standard model gauge group,
the massive components that can affect the running of the coupling constants
are: the vector multiplets in 
( (\rep{2},\rep{3})$_{-5/3}$ + h.c.\ )  [$V$], 
( (\rep{1},\rep{1})$_2$ + h.c.\ )       [$V_1$],
( (\rep{1},\repC{3})$_{-4/3}$ + h.c.\ ) [$V_3$], and
( (\rep{2},\rep{3})$_{1/3}$ + h.c.\ )   [$V_6$]
representations; and the chiral multiplets in
(\rep{3},\rep{1})$_{0}$ [$A_3$], 
(\rep{1},\rep{8})$_{0}$ [$A_8$], 
( (\rep{3},\rep{1})$_{2}$ + h.c.\ )     [$D_3$], 
( (\rep{2},\rep{3})$_{1/3}$ + h.c.\ )   [$D_6$], 
( (\rep{1},\rep{6})$_{-4/3}$ + h.c.\ )  [$D_6'$], 
( (\rep{1},\rep{1})$_2$ + h.c.\ )       [$B_1$],
( (\rep{1},\repC{3})$_{-4/3}$ + h.c.\ ) [$B_3$],
( (\rep{2},\rep{3})$_{1/3}$ + h.c.\ )   [$B_6$],
( (\rep{2},\rep{1})$_{1}$ + h.c.\ )     [$C_2$], and
( (\rep{1},\rep{3})$_{-2/3}$ + h.c.\ )  [$C_3$]
representations where the labels we will use for them are indicated within
the square braces.   In addition, there are two triplets and a doublet
($H_3$, $T_3$, $T_2$) coming from the two \rep{10}'s.
The $A$'s come from the \rep{45}, the $D$'s from the 
\rep{54}, the $B$'s are a combination of the \rep{45} and
the \rep{16} + \repC{16}, and the $C$'s are from the \rep{16} + \repC{16}. 
When the unification condition is imposed, we find then that:
\bea
\Bigl( \frac{\Mpd}{M_{H_3}^{(0)}} \Bigr) &=&
\Bigl( \frac{ \Msusy }{ \Mz } \Bigr)^{5/6}
\Bigl( \frac{ M_{V_1} M^3_{V_3} }{ M^4_{V_6} } \Bigr)
\Bigl( \frac{ M_{A_3} }{ M_{A_8} } \Bigr)^{5/2}  \nonumber \\ 
&\times&
\Bigl( \frac{ M^4_{B_6} }{ M_{B_1} M^3_{B_3} } \Bigr)^{1/2}
\Bigl( \frac{ M_{C_2} }{ M_{C_3} } \Bigr)
\Bigl( \frac{ M^7_{D_3} M^4_{D_6} }{ M^{11}_{D_6'} } \Bigr)^{1/2}.
\label{eq:Mpd3}
\eea

In the one-step unification picture, the threshold corrections are expected
to be small.  In certain SO(10) models, on the other hand, the necessity
to not badly destabilize the DW form of VEV is satisfied by weakly coupling 
the two gauge breaking sectors which in turn leads to introduction of
of intermediate mass scales\cite{Babu:1995,Hisano:1994}.  
In these models, the massive states
$B_3$ and $B_6$ are of intermediate scale while all the other states
(including $B_1$) are at the GUT scale.
Thus for these models, the threshold effect in fact enhances the proton 
decay rate.  

Now we turn to the possible solutions to the problems raised here.
The results above point out how the proton decay rate is intimately related 
to the threshold effects in these SO(10) models.  
Bringing these models into agreement with experiment hence require large
threshold effects.  One is then faced with having to explain two
large effects: (1) why the value of 
$\Mpd = M_{T_3} M_{H_3} / M_{T_2}$ is much larger than $\Mgut$, and
(2) why the threshold effects (the right hand side of Eq.~(\ref{eq:Mpd3})
is also large, in the right direction, and of the correct size 
to explain the ratio $\Mpd/M_{H_3}^{(0)}$.

One possible way to explain both of these effects is simply to adjust
the values of the threshold masses\cite{foot2}.
If we take for granted that item (1) is
solved by adjusting $\Mpd$ to be large enough to satisfy Eq.~\ref{eq:mpdlim}
for $\tan \beta \simeq 50$, then item (2) can be accomplished by a threshold
effect to produce a shift in $\alpha_3(\Mz)$ $\gtrsim 13\%$, or 
$\gtrsim 5 $ std.\ as can be seen from Fig.~1.  
If there are intermediate mass states as discussed above, 
it provides a threshold effect in the wrong direction, 
and the remnant threshold effect must then be $\gtrsim 40 \%$, or 
$\gtrsim 16 $ std.\  in order
to bring the ratio $\Mpd/M_{H_3}^{(0)}$ into agreement with experiment.
These are to be contrasted with the ``successful''
prediction of $\alpha_3(\Mz)$ (within $\lesssim 4 \%$ or $\lesssim 2 $ std.) 
within the framework
of supersymmetric grand unification where thresholds are neglected.
Thus the naturalness of grand unification in such models is eroded.

Perhaps a more natural solution would be to introduce a mechanism which
suppresses proton decay in a different way, so that it is not strongly
dependent on threshold masses.  For example, with 
three \rep{10}'s, three \rep{45}'s and 
two \rep{54}'s, Babu and Barr have shown it is
possible to strongly suppress proton decay\cite{Babu:1993}.
Such models then require a much larger particle spectrum.

In conclusion, we have shown that the proton decay constraint plays an important
role for minimal SO(10) models where the largeness of $\tan \beta$ requires
that there be a mechanism to suppress the Higgsino mediated proton decay.
We have shown that the mass matrix mixing among the two pairs of
superheavy triplets that occur naturally in the Dimopoulos-Wilczek mechanism
does not in itself give rise to such a suppression, since the proton decay rate 
is constrained by the unification condition in exactly the same manner as 
without such a mixing.  We have also shown that in those
models where intermediate mass states exist, proton decay is in fact
further enhanced, making the proton decay constraint even more difficult to
satisfy.

This work was supported in part by the National Science Foundation
under grant number PHY-9411543.

\begin{thebibliography}{10}

\bibitem[\dagger]{shinichi}
Email: Urano@phys.tamu.edu (internet).

\bibitem[\ast]{dick}
Email: Arnowitt@phys.tamu.edu (internet).

\bibitem{Dimopoulos:1981inp}
S. Dimopoulos and F. Wilczek,  in {\em The Unity of the Fundamental
  Interactions}, Proceedings of the 19th Course of the International School of
  Subnuclear Physics, Erice, Italy, 1981, edited by A. Zichichi (Plenum Press,
  New York, 1983).

\bibitem{Srednicki:1982}
M. Srednicki, Nucl. Phys. {\bf B202},  327  (1982).

\bibitem{Dienes:1996a}
K.~R. Dienes, A.~E. Faraggi, and J. March-Russell, Nucl. Phys. {\bf B467},  44
  (1996).

\bibitem{Hisano:1994}
J. Hisano, H. Murayama, and T. Yanagida, Phys. Rev. D {\bf 49},  4966  (1994).

\bibitem{Babu:1995}
K.~S. Babu and S.~M. Barr, Phys. Rev. D {\bf 51},  2463  (1995).

\bibitem{foot1}
See, for example, Appendix 1 of \cite{Lucas:1996} and \cite{Babu:1993} where
  such a mixing is presented as a mechanism to suppress proton decay.

\bibitem{Nath:1985}
P. Nath, A.~H. Chamseddine, and R. Arnowitt, Phys. Rev. D {\bf 32},  2348
  (1985).

\bibitem{Arnowitt:1994}
R. Arnowitt and P. Nath, Phys. Rev. D {\bf 49},  1479  (1994).

\bibitem{Jones:1982}
D.~R.~T. Jones, Phys. Rev. D {\bf 25},  581  (1982).

\bibitem{Antoniadis:1982}
I. Antoniadis, C. Kounnas, and K. Tamvakis, Phys. Lett. B {\bf 119},  377
  (1982).

\bibitem{Bagger:1995}
J. Bagger, K. Matchev, and D. Pierce, Phys. Lett. B {\bf 348},  443  (1995).

\bibitem{Ring:1995}
D. Ring, S. Urano, and R. Arnowitt, Phys. Rev. D {\bf 52},  6623  (1995).

\bibitem{foot2}
This is essentially the approach taken in the model examined in
  \cite{Lucas:1996} which has a large number of thresholds spanning a range of
  $10^{13} - 10^{20}$~GeV. Though predictivity and naturalness is thus
  diminished, this model arranges these thresholds in such a way as to
  substantially lower the prediction of $\alpha_3(\Mz)$ for a wide region of
  its parameter space, thus partially explaining item (2).

\bibitem{Babu:1993}
K.~S. Babu and S.~M. Barr, Phys. Rev. D {\bf 48},  5354  (1993).

\bibitem{Lucas:1996}
V. Lucas and S. Raby, preprint: OHSTPY-HEP-T-96-030, hep-ph/9610293.

\bibitem{Langacker:dpf96}
P. Langacker, talk given at DPF96, Minneapolis, Aug. 1996.

\end{thebibliography}
\end{document}